\documentclass[10pt]{article}
\usepackage[OE]{express}
\usepackage{graphicx,subfigure}

\begin{document}

%%%%% Titles %%%%%%%

%\title{Unexpected modal noise in a diffraction-limited photonic spectrograph}
%\title{Unexpected modal noise in an integrated photonic lantern fed diffraction-limited spectrograph}
\title{Modal noise in an integrated photonic lantern fed diffraction-limited spectrograph}

%%%%% Authors %%%%%%%

\author{N. Cvetojevic,\authormark{*,1,2,3} N. Jovanovic,\authormark{4,5} S.Gross,\authormark{1,5} B. Norris,\authormark{3} \\ I. Spaleniak,\authormark{5} C. Schwab,\authormark{2,5}  M. J. Withford,\authormark{1,5}, M. Ireland,\authormark{6} \\ P. Tuthill,\authormark{3} O. Guyon,\authormark{4,7,8} F. Martinache,\authormark{9} and \\ J. S. Lawrence\authormark{2,5}}

\address{\authormark{1}Centre for Ultrahigh bandwidth Devices for Optical Systems (CUDOS), Australia\\
\authormark{2}The Australian Astronomical Observatory  (AAO), 105 Delhi Rd, North Ryde, NSW 2113, Australia\\
\authormark{3}Sydney Institute for Astronomy (SIfA), School of Physics, University of Sydney, NSW 2006, Australia\\
\authormark{4}Subaru Telescope, National Astronomical Observatory of Japan, National Institutes of Natural Sciences (NINS), 650 North A'Ohoku Place, Hilo, HI, 96720, U.S.A.\\
\authormark{5}MQ Photonics Research Centre, Dept. of Physics and Astronomy, Macquarie University, NSW, Australia\\
\authormark{6}Research School of Astronomy \& Astrophysics, Australian National University,  ACT, Australia\\
\authormark{7}Steward Observatory \& College of Optical Sciences, University of Arizona, Tucson, AZ, 85721, U.S.A.\\
\authormark{8}Astrobiology Center of NINS, 2-21-1, Osawa, Mitaka, Tokyo, 181-8588, Japan\\
\authormark{9}Laboratoire Lagrange, Universit\'{e} C\^{o}te d'Azur , Observatoire de la C\^{o}te d'Azur, CNRS, Parc Valrose, B\^{a}t. H. FIZEAU, 06108 Nice, France\\
}

\email{\authormark{*}nick.cvetojevic@gmail.com} %% email address is required

%%%%%%%%%%%%%%%%%%% Abstract and OCIS codes %%%%%%%%%%%%%%%%
%% [use \begin{abstract*}...\end{abstract*} if exempt from copyright] ????

\begin{abstract}
In an attempt to develop a streamlined astrophotonic instrument, we demonstrate the realization of an all-photonic device capable of both multimode to single mode conversion and spectral dispersion on an 8-m class telescope with efficient coupling. The device was a monolithic photonic spectrograph which combined an integrated photonic lantern, and an efficient arrayed waveguide grating device. During on-sky testing, we discovered a previously unreported type of noise that made spectral extraction and calibration extremely difficult. The source of the noise was traced to a wavelength-dependent loss mechanism between the feed fiber's multimode near-field pattern, and the modal acceptance profile of the integrated photonic lantern. Extensive modeling of the photonic components replicates the wavelength-dependent loss, and demonstrates an identical effect on the final spectral output. We outline that this could be mitigated by directly injecting into the integrated photonic lantern. 
\end{abstract}

\ocis{(120.6200) Spectrometers and spectroscopic instrumentation; (130.2755) Glass waveguides; (350.1260) Astronomical optics; (130.3120) Integrated optics devices.}
%For a complete list of OCIS codes, visit: https://www.osapublishing.org/oe/submit/ocis/

%%%%%%%%%%%%%%%%%%%%%%% References %%%%%%%%%%%%%%%%%%%%%%%%%

%%%%%%%%%%%%%%%%%%%%%%%%%%%%%%%%%%%%%%%%%%%%%%%%%%%%%%%%%%%%%%%%%%%%%%%%%%%
%   Body of Paper
%%%%%%%%%%%%%%%%%%%%%%%%%%%%%%%%%%%%%%%%%%%%%%%%%%%%%%%%%%%%%%%%%%%%%%%%%%%

%%%%%%%  Introduction  %%%%%%%%%
\section{Introduction}

With the dawn of the extremely large telescope (ELT) age just over the horizon, the astronomical community is becoming quickly confronted by the reality of the scaling cost of state-of-the-art spectrographic instrumentation. The next generation of spectrographs for the ELTs (such as GMACS~\cite{GMACS}, GCLEF~\cite{GCLEF} among others) require considerably larger investments, in both size and cost than those of previous generations. They also drive the requirements on the optical components to the very limit of what is currently achievable. This is because of the scaling law for instruments as a function of the telescope aperture, which do not operate in the diffraction-limited regime (well established in the instrumentation community and originally discussed in an astrophotonic context in ~\cite{josslaw} and more recently treated in~\cite{jov2016}). It is perhaps then unsurprising that recently there has been a flurry of development and interest in diffraction-limited (single-mode) alternatives, which not only break this fundamental scaling law, but also provide new advantages (can remove modal-noise, for example)~\cite{jov2016,rains2016,crepp16}.

Despite the potential advantages, diffraction-limited spectroscopy has a major hurdle to overcome: how to efficiently inject light that has been perturbed by the atmosphere and convert it to a diffraction-limited state for spectroscopy. If diffraction-limited light is to be fed to the spectrograph, it is clear that a single-mode fiber (SMF) will be used to transport the light. SMFs only allow the propagation of the fundamental mode within the fiber, which is characterized by a Gaussian intensity profile and a flat phase front. This feature provides spatial filtering at the fiber input by rejecting all higher order modes, which makes the output beam profile temporarily invariant in shape, and is highly sought-after for high-resolution spectroscopy and certain interferometric applications. This comes at a great cost in terms of coupling efficiency, as any higher order modes in the focal plane, typical of a seeing-limited or an adaptive optics (AO)-corrected beam, are filtered out. Thus, to achieve high coupling, the light must have a flat wavefront. In addition, even if the wavefront is flattened, there is also a mismatch between the Airy pattern at the telescope focus and Gaussian mode of the fiber, which is commonly left unaddressed and places an upper-limit to the maximum coupling efficiency of $\sim80\%$ (assuming an unobstructed pupil)~\cite{Shaklan1988}. Thus, the typical coupling values achieved without any correction are of the order of a percent, while with AO correction, it typically increases to tens of percent~\cite{Coude2000,woillez2003}. Even at these levels of coupling, it is hard to justify their use except for in the most specific of circumstances.  

Photonic Lanterns (PLs) allow for a translation of a multimoded input into a series of single-moded outputs (and vice versa)~\cite{saval2013}. The detailed operation of these devices is described later on, but it is worth emphasizing that the possibility of converting a beam from seeing to diffraction-limited has been implemented into astronomical instruments~\cite{jbh2011,Schwab2012}. Furthermore, with the original lanterns initially developed using optical fibers, a number of groups have explored the manufacture and use of waveguide-based lanterns, entirely written as a $3$-D circuit in a glass chip~\cite{Iza1,Iza2,RthompsonPL}. These integrated photonic lanterns (IPLs) allow for significant miniaturization and simplification, especially when a large number of modes are required. Promisingly, IPLs have recently been tested on telescope~\cite{HarrisOnSky, MacLachlan}, demonstrating an improved throughput ($40$--$50\%$) when an AO system is used (without feeding an instrument).  

We aim to take one step further by combining an IPL with an integrated photonic spectrograph (IPS) to form a monolithic fiber-fed device and test it on a large telescope. The devices we fabricated were designed to capture seeing-limited light at the telescope focal plane, convert it into multiple single-mode (SM) channels, and spectrally disperse the channels simultaneously, all on a single detector. This small device has the throughput advantages of working in the seeing-limited regime, while providing all the benefits of a diffraction-limited photonic spectrograph. This breakthrough is a natural evolution for astrophotonic instrumentation, bringing them more in line with modern research telescopes, and broadening the niche where such technology can potentially be utilized.

Although this is an obvious evolution for this technology, we will elucidate some shortcomings of design, especially when interfacing multimode fibers (MMFs) and multimode waveguides (MMWGs), which have the potential to cause an unexpected and uncalibratable form of wavelength dependent loss, which manifests as high frequency noise across the spectrum. The purpose of this study is to showcase this subtle loss mechanism, explain it, and outline mitigation strategies for the future. If left unchecked, this new form of noise makes acquired on-sky spectra uncalibratable, meaning its discovery is topical and will be of interest to the broader astrophotonic community considering integrating such components into instrumentation. In Section~\ref{sec:Background} we provide a brief overview of the photonic components used, and present the concept of our integrated chip. Section~\ref{sec:ExperimentSetup} outlines the fabrication and characterization setups used, with the results presented in Section~\ref{sec:Labresults}. Finally, Section \ref{sec:ModalNoise} discusses the nature and cause of the unique type of modal noise discovered in our system. We further present beam-propagation simulations that outline and describe the root cause of the noise and demonstrate mitigation techniques we used to overcome it.

%%%%%%%	  Background  	%%%%%%%%%

\section{Instrument concept and design} \label{sec:Background}
The aim of this section is to offer deeper insights into the operation of both IPLs and IPSs, which underpin the instrument concept presented below.  

\subsection{The integrated photonic lantern}
Photonic lanterns can efficiently convert a multimode (MM) input to numerous diffraction-limited outputs if the number of single-mode waveguides (SMWGs) matches or exceeds the number of modes excited in the MM end of the device, and if the transition is adiabatic~\cite{saval2013}. The device was originally created by tapering down a capillary tube filled with SMFs until they formed one multimode fiber (MMF) \cite{SergioPLOrigional}. However, if a large number of modes are present, as is the case, for example, if one moves to shorter wavelengths (i.e. the visible), then the number of SM channels can grow rapidly. This places tough demands on the fabrication process for fiber based devices, and if successfully fabricated, the high fiber count makes the device unwieldy to work with. This was recognized at the outset as a limitation \cite{SergioPLOrigional}, with a solution being developed by Tim Birks et. al. ~\cite{birks,birks2}, who proposed using a single multicore fiber (MCF) to create the lantern instead. This approach greatly simplified the device, and subsequent handling requirements. However, because the cores at the MCF output are arranged in a $2$D array, it still requires some form of re-formating to bring it to a $1$D array to form a spectrograph slit.

In an alternative approach, ultrafast laser inscription (ULI) was used to fabricate an integrated device around the same time in a glass chip, achieving the ultimate level of miniaturization ~\cite{RthompsonPL}. This preliminary result used the multi-scan technique whereby the writing laser conducts numerous overlapping passes to build up each SM, and ultimately the multi-mode waveguide (MMWG), which make up the device. Devices fabricated by this method have been tested on-sky~\cite{HarrisOnSky}. This work successfully demonstrated efficient coupling behind an AO system (CANARY AO at the William Herschel Telescope) achieving throughputs of $\sim 20\%$ under particular conditions using a $36-$port lantern. The SM outputs of the lantern did not ultimately feed a spectrograph, but did demonstrate the feasibility of improving telescope coupling for potential diffraction-limited science. 

Following on from this work, an approach was used to fabricate the first IPLs that ensured the laser did not overlap any modified regions, which offers more reproducible waveguide properties. The technique was first used to fabricate MMWGs, which are required at the input of an IPL~\cite{jov2012}. As a second step, these guides were incorporated into a very efficient IPL device~\cite{Iza1}. These devices offered near lossless transitions at high focal ratios ($>8$). As a final step, a single device with a MMWG, a lantern, and fiber Bragg gratings for spectral filtering, was also realized~\cite{Iza2}. Devices fabricated by this latter technique underpin the components utilized in this study.

\begin{figure}[ht!]
	\centering\includegraphics[width=0.83\linewidth]{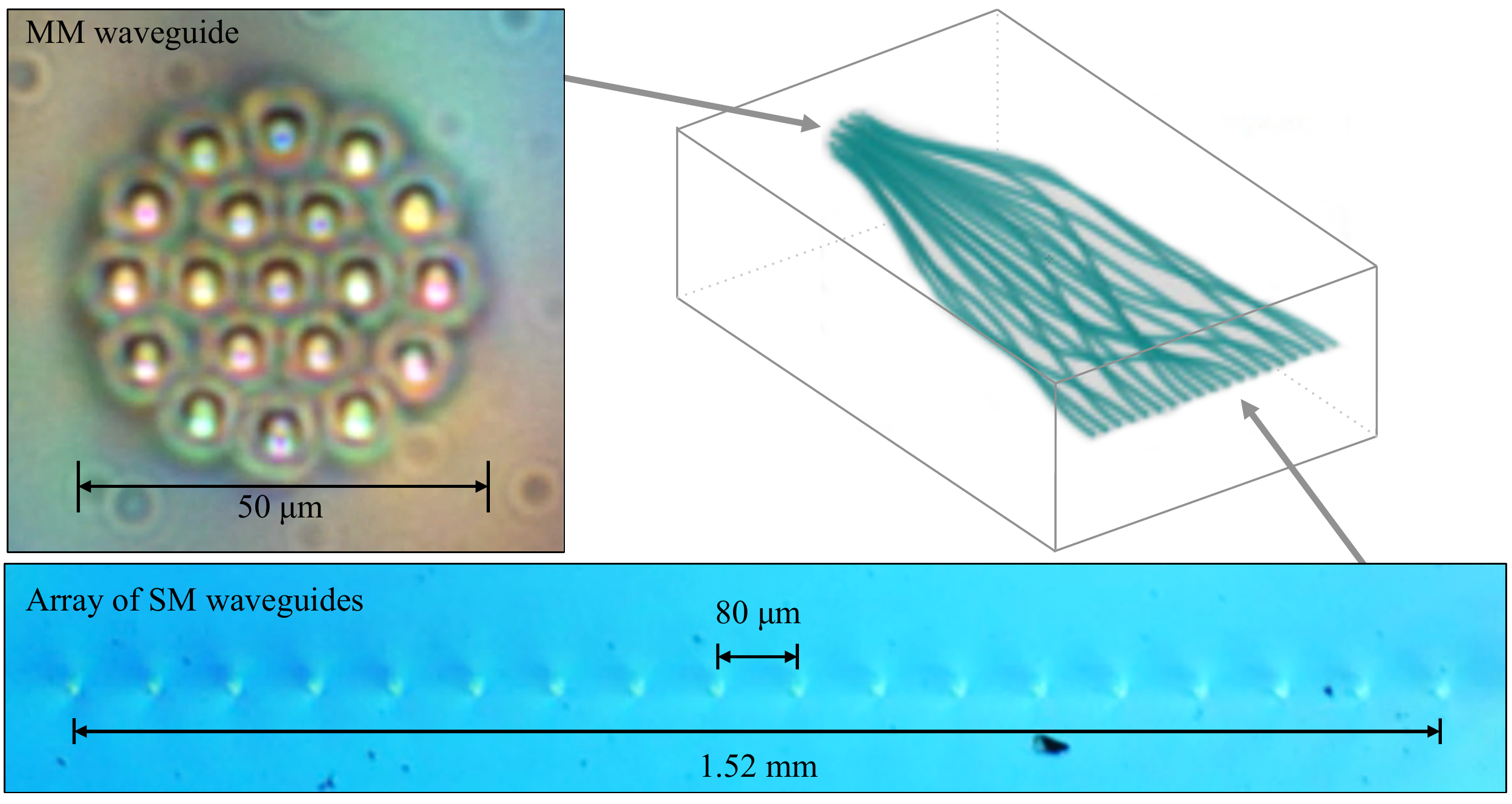}
    \caption{The IPL consists of multiple continuous refractive index modifications which act as waveguides inscribed in a transparent glass substrate (top right). Micrographs of the input and output array of guides for a fabricated device are shown.}
    \label{fig:IPLSchem}
\end{figure}

The IPL devices presented in this work were inscribed in EAGLE2000 alumino-borosilicate glass, using the aforementioned ULI fabrication method, that created positive refractive index changes in the bulk glass by focusing a Ti:sapphire laser (FEMTOSOURCE XL 500, Femtolasers GmbH). The WGs were inscribed with a laser power of $35$~nJ at a $5.1$~MHz rep rate, and a translation velocity of $750$mm/minute. Because of the high repetition rates used, the waveguides were written in the cumulative heating regime \cite{CumHeating}. Two lantern architectures were fabricated, a $1\times37$ and $1\times19$ port lantern, with the latter depicted in Fig.~\ref{fig:IPLSchem}. The IPLs had a $2$~mm long MMWG section created by making a closely spaced circular array of positive refractive index modifications (i.e. SMWGs) such that they form a single larger waveguide (i.e. the waveguides are close enough that there is significant overlap between the modes creating MMWG)~\cite{jov2012}. The MMWG in the $1\times19$ IPL consisted of three rings of refractive index modifications (SMWGS), and had an effective diameter of 50~\textmu m. This meant the MMWG was well matched to standard MMFs in size.

The MMWG underwent an adiabatic transition to a $2$-D array of $19$ individual SMWGs. This was achieved by gradually increasing the separation between the SMWG tracks in a cosine arc profile over the $20$~mm span of the device, as determined in a previous study~\cite{Iza1}. At the end of this transition, the light was confined to the SMWGs, and each track was considered as an isolated channel. The $2$-D array of SMWGs is incompatible with the planar devices we wish to feed, thus a remapping region was added. This region transformed the $2$-D array into a uniformly spaced $1$-D array, with a pitch of 80~\textmu m. The linear array had an overall span of $1.52$~mm and is shown in Fig.~\ref{fig:IPLSchem}.

To aid in rapid prototyping and characterization, a straight MMWG and SMWG were written with the same laser parameters adjacent to the lantern in each glass chip to calibrate throughput and coupling measurements. Additional visible markings (that do not guide light) were inscribed to aid in identifying the structures when imaging and aligning. Multiple duplicate chips were fabricated such that any unfavorable variation in the fabrication procedure could be identified during chip characterization, and rejected. After fabrication, the waveguide ends of the chips were polished to optical quality on the entry-and-exit facets, resulting in a final chip length of $30$~mm.

\subsection{The integrated photonic spectrograph} \label{Sec:introAWG}
Arrayed Waveguide Gratings (AWGs) are one of the more mature photonic technologies to be explored for astronomical spectroscopy~\cite{cvet2009, cvet2012a}. This form of on-chip spectrograph requires the injection of light from a SMF, with the resulting spectra from all grating orders superimposed on top of one another at the chip output. The devices characterized in~\cite{cvet2009, cvet2012a} were commercial-grade AWGs, optimized for operation around $1550$~nm, with a free-spectral range (FSR) of $50$~nm, and a resolving power of R~=~$7000$. In order to obtain broader wavelength coverage, the orders of the AWG have to be separated with cross-dispersion via a prism or grating. This approach has been used to collect spectra across the entire H-band ($1490$--$1800$~nm) on-sky~\cite{cvet2009, cvet2012b}. The peak internal AWG throughput was $77\%$.

AWGs are a lithographically fabricated planar photonic device that consist of three key sections: the waveguide array that creates the wavelength dispersion, and the input and output free-propagation zones (FPZs), which act like the collimating and focusing optics in a classical spectrograph. A schematic of a device is shown in Fig.~\ref{fig:AWGSchimatic}. The FPZs are slab waveguides that guide light in a SM in the vertical plane but allow the light to diverge in the substrate plain. Light enters the input FPZ typically from a SMF. At the opposite end of the FPZ, an array of SMWGs are uniformly positioned in such a way that they track the curvature of the wavefront at that location, thus the light enters the array along a single phasefront of the diffracting input, referred to as the Rowland circle. The waveguides in the array route the light to the output FPZ (mirrored version of the input FPZ), with each waveguide being incrementally longer than the adjacent one by a designed amount. This creates a fixed optical delay between each waveguide in the array so when the beams are combined in the output FPZ, interference takes place, which directs light of a given wavelength in a specific direction, forming a spectrum at the output of the FPZ. 

The AWG devices were commercially available, off-the-shelf items. They were fabricated using Silica-on-Silicon (SoS) lithography procedures, onto a larger wafer that contains multiple devices, and were subsequently diced up into individual components. The AWGs used in this work were slightly modified, as detailed in~\cite{cvet2012a}.

\subsection{Combining technologies} \label{Sec:combo}

With the PL capable of producing multiple SM output channels, AWGs were optimized early on so that signals from multiple SM inputs could be injected simultaneously into a single AWG chip and recovered via cross-dispersion~\cite{cvet2012a}. As was the case for separating the orders of a single input channel, the spectrum of each SM input could also be disentangled with the aid of cross-dispersion, and collected on the detector. This concept was tested on the Anglo-Australian Telescope (AAT), using a fiber-based PL developed originally for the GNOSIS instrument~\cite{trinh2013}, demonstrating that compact SM spectrographs could be used even on seeing-limited telescopes. However, due to the mismatch between the number of SMFs that could be interfaced to an AWG ($\sim12$) and the number of modes in a typical seeing-limited spot on the AAT (hundreds), the overall throughput of the system was low ($<5\%$). 

\begin{figure}[ht!]
    \centering\includegraphics[width=0.83\linewidth]{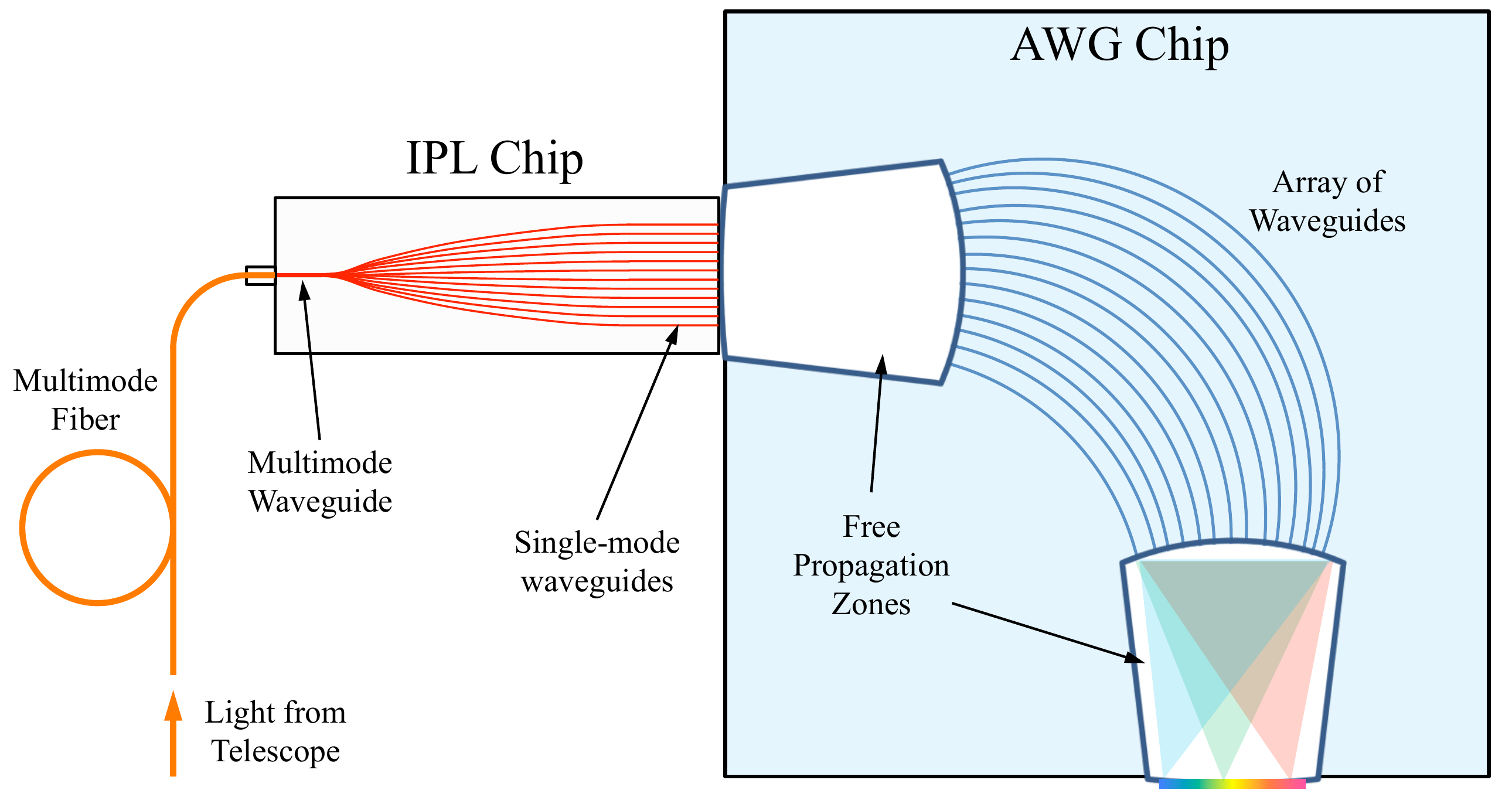}
    \caption{Schematic showing the photonic components of the final spectrograph assembly. The light from the telescope was coupled to a $2$~m long MMF, which was bonded to the MMWG input of the IPL chip. The lantern transitioned and remapped into a linear array of individual SMWGs. These outputs were used to directly inject into the AWG, which forms the spectra of each SM channel at the chip output.}
    \label{fig:AWGSchimatic}
\end{figure}

To alleviate this mismatch, and thus increase the end-to-end efficiency of the instrument, the prototype used at the AAT was improved in two ways. Firstly, the fiber-based PL was replaced with an IPL as outlined above. With the IPL fabricated in a glass block, it was relatively easy to bond directly to the AWG chip, requiring no further fiber splicing, reducing the preparation time, and minimizing loss. With the IPL and AWG bonded, the MM input of the IPL was aligned and bonded to a MMF, into which telescope light could be coupled at the focal plane. To match the IPL MMWG section, a core size of 50~\textmu m was used for the fiber. As part of the testing, we used two versions of the MMF, a standard $0.22$~NA, and a lower index-contrast fiber with a $0.12$~NA, which better matched the modal capacity of the lantern.  A schematic of the device is shown in Fig.~\ref{fig:AWGSchimatic}. With the entire integrated photonic spectrograph (IPS) unit having a physical footprint of a few centimeters squared, it succeeds in miniaturizing both the modal conversion and spectral dispersion into a compact unit, while maintaining a MMF feed standard for many telescopes. 

%%%%%%%	  Experiments  	%%%%%%%%%
\section{Experiments}\label{sec:ExperimentSetup}
This section outlines the fabrication and characterization of the device outlined in the previous section, and its subsequent integration into a complete spectrograph.  

%%%%%%%
\subsection{Photonic device characterization} \label{sec:devicecharacterizationsetup}
\begin{figure}[ht!]
	\centering\includegraphics[width=0.99\linewidth]{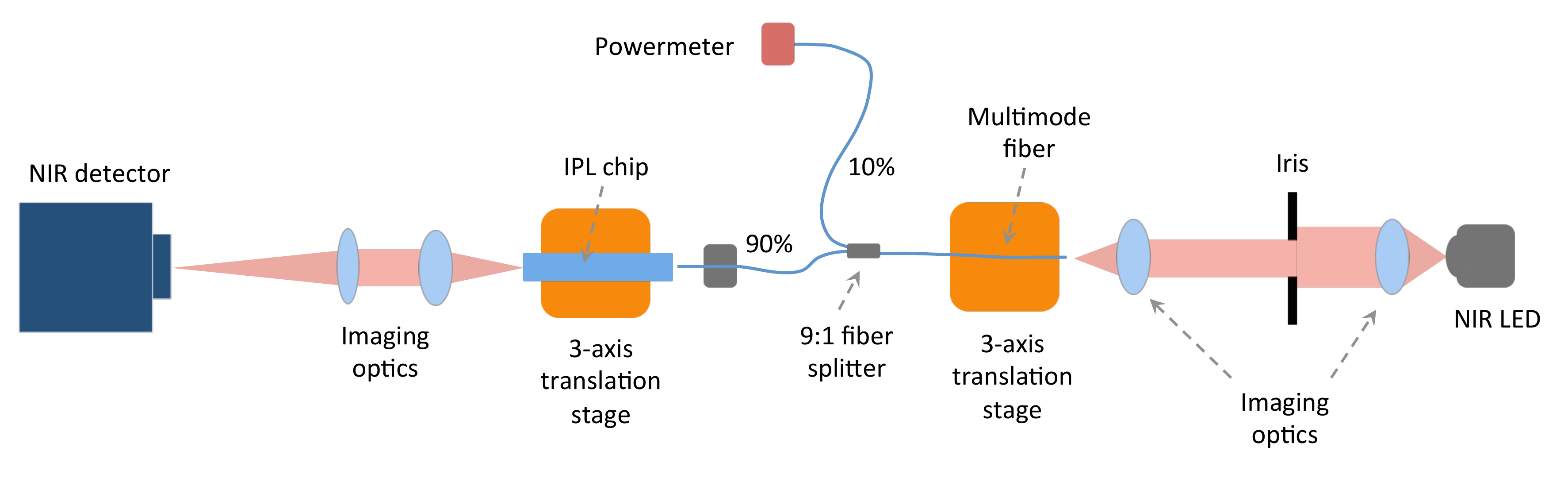}
    \caption{The individual photonic components of the spectrograph were characterized by imaging the chip output, and calibrated using a powermeter. The injection comprised of a re-imaged NIR LED with an iris for controlling the injected f$/\#$ (inversely proportional to NA).}
    \label{fig:Labsetup}
\end{figure}

The experimental setup was designed to measure the overall throughput of the IPLs as a function of the focal ratio of injection (f/$\#$), and is presented in Fig.~\ref{fig:Labsetup}. The injection portion of the setup consisted of a broadband NIR LED source re-imaged onto a 50~\textmu m core MMF for the $1\times19$ (and 100~\textmu m core for the $1\times36$) using two achromatic NIR lenses. Between the lenses, in the collimated beam, an adjustable iris was placed to control the injected f/$\#$ by stopping down the beam. The fiber output was cleaved and placed on a stationary bracket. The IPL was placed on a translation stage, such that its output was imaged by a NIR detector. The detector was power calibrated using a pickoff to a power meter which provided simultaneous power measurements post-coupling. The block of glass was translated to align each device to MMF in use, and the output of the MMWG and the SMWGs of the PLs were measured.  The MMWG was used to calibrate the relative throughputs of the WG section and the PL independently, and observe how the injected f/$\#$ affected them. This was important to ensure that the MMWG was not limiting the accepted f/$\#$ but rather only the lantern itself.  

%%%%%%%
\subsection{Device bonding and assembly}
A MMF was aligned and bonded to the IPL using a multi-axis piezo actuated translation stage. The MMF was cleaved and placed onto the translation stage, which could align in the X, Y and Z directions as well as pitch, yaw, and roll. The roll axis was useful for when it came to aligning the IPL to the AWG device. A ferrule was placed over the fiber end-face to provide a larger surface of contact, and increase the strength of adhesion of the MMF to the IPL. The MMF had broadband light injected into it, and the IPL output imaged using a Xenics Xeva-1.7 640 InGaAs detector, such that all the IPL outputs were visible (and the output flux measurable) at the same time. The fiber was then aligned by maximizing the coupling to all individual channels, while also maintaining the overall uniformity and total flux. It was then bonded using UV curing epoxy. This was done for both the $1\times19$ and $1\times36$ devices using 50~\textmu m and 100~\textmu m core fiber, respectively. Both fibers had a numerical aperture (NA) of $0.12$. After bonding, the lanterns were characterized using the setups mentioned in the previous subsection. 

\begin{figure}[ht!]
    \centering\includegraphics[width=0.83\linewidth]{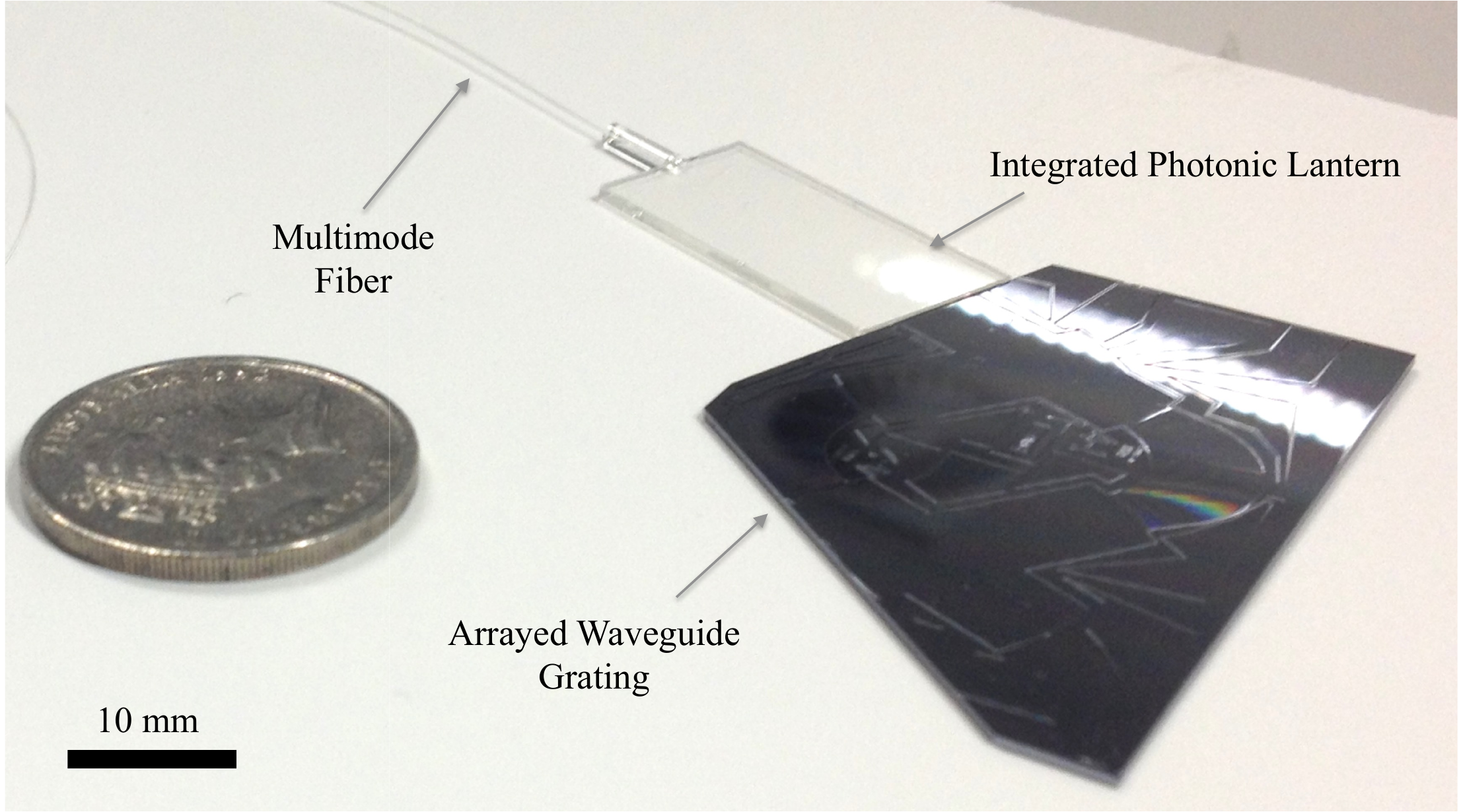}
    \caption{The completed on-sky device after bonding is shown with a coin (23.6~mm diameter) to provide scale.}
    \label{fig:AWGassembly}
\end{figure}

After characterization, the IPL module was bonded to the AWG chips using a similar method. The IPL was placed onto the translation stage, with the AWG chip stationary under a vision system for precise alignment. The AWG output was imaged using the NIR detector to observe the full chip output. Due to the wavelength dispersion of the chip, a narrow-band, tunable NIR laser source was used that could provide a monochromatic source to avoid overlapping light from multiple orders. In this way, all ports of the lantern could be imaged through the AWG, while wavelength tuning of the laser allowed rapid confirmation of the alignment and presence of the other wavelengths. The metric used was output power of the individual guides to ensure optimal alignment, with additional observation of image quality to ensure centering on the FPZ. The IPL was aligned and bonded similarly as above, this time with careful attention paid to the chip rotation. After the curing process, the chip was placed into a metal protective housing for shipping. The final bonded device (pre-packaging) is shown in Fig.~\ref{fig:AWGassembly}. 

%%%%%%%
\subsection{The SCExAO instrument}
The photonic spectrograph assembly was tested at the $8$~m Subaru Telescope on Mauna Kea, using the Subaru Coronagraphic Extreme Adaptive Optics (SCExAO) instrument. The SCExAO instrument is a platform that, when commissioned, will offer extreme AO correction, achieving up to $90\%$ Strehl ratios in the H-band, and offer numerous coronagraphs and interferometers optimized for imaging faint companions, such as planets close to their host stars. For a full description of the instrument please refer to~\cite{jov2015}. Here, we offer an overview of the aspects that are relevant to this work.

During the period of the tests carried out in this work ($2014$), the extreme AO correction was not operational. AO correction was provided by the AO188 instrument~\cite{min2010}. AO188 is upstream of SCExAO and facilitates an initial correction, which SCExAO will eventually improve upon. The results in this section were obtained with Strehl ratios of $30$--$40\%$ in the H-band.  

\begin{figure}[ht!]
	\centering\includegraphics[width=0.99\linewidth]{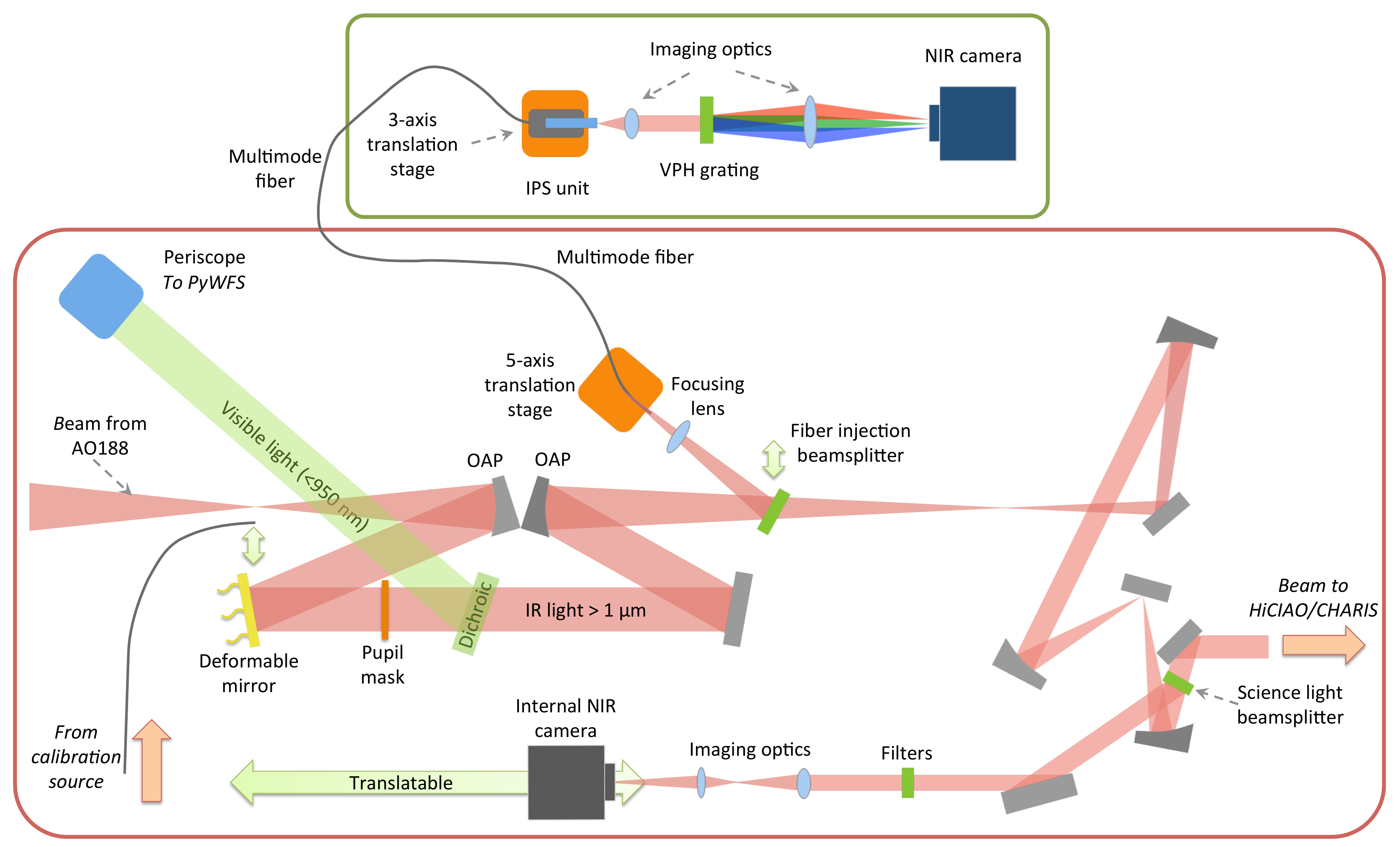}
    \caption{A schematic of the SCExAO instrument in $2014$ when the experiments were carried out. Dual head green arrows indicate that a given optic can be translated in/out of, or along the beam. Orange arrows indicate light entering or leaving the designated bench at that location. OAP-off axis parabolic mirror, PyWFS-pyramid wavefront sensor. The IPS was fed by a MMF, which was injected into by a fibre-injection module built into SCExAO. The AWG output was spectrally dispersed with a VPH cross-disperser and re-imaging optics on a small breadboard to the side of SCExAO. The cross-dispersion separated the initially overlapping spectra from each SM lantern output, as well as the multiple grating orders of the AWG device.}
    \label{fig:scexao}
\end{figure}

The corrected light was injected into the MMF using a fiber injection module inside SCExAO, shown in Fig.~\ref{fig:scexao}~\cite{jov2014}. The fiber injection module allows for precise active alignment of the MMF with the corrected focal spot. An off axis parabolic mirror was used to focus the beam. A small aspheric lens was placed close to the focus to modify the focal ratio of the beam as required for efficient injection. By varying the distance between the lens and the off axis parabola, the focal ratio was adjusted. For a comprehensive explanation of the fiber-injection system on SCExAO see~\cite{jov2014,jov2017}. 

The all-photonic spectrograph was integrated into a small bulk-optic instrument which cross-dispersed the output of the AWG and imaged it onto a detector. Cross-dispersion was achieved by using an off-the-shelf VPH grating. Achromatic lenses were used to provide the adequate magnification ($\times10$) and sampling on the InGaAs detector (same as used in laboratory experiments). The instrument is shown in Fig.~\ref{fig:scexao}.

%%%%%%%	  Lab Results  	%%%%%%%%%
\section{Laboratory results}\label{sec:Labresults}
Before attempting to utilize the photonic instrument on sky for science, it was important to fully characterize all aspects of it.

\subsection{Photonic device characterization}
\begin{figure} [ht!]
	\centering\includegraphics[width=0.95\linewidth]{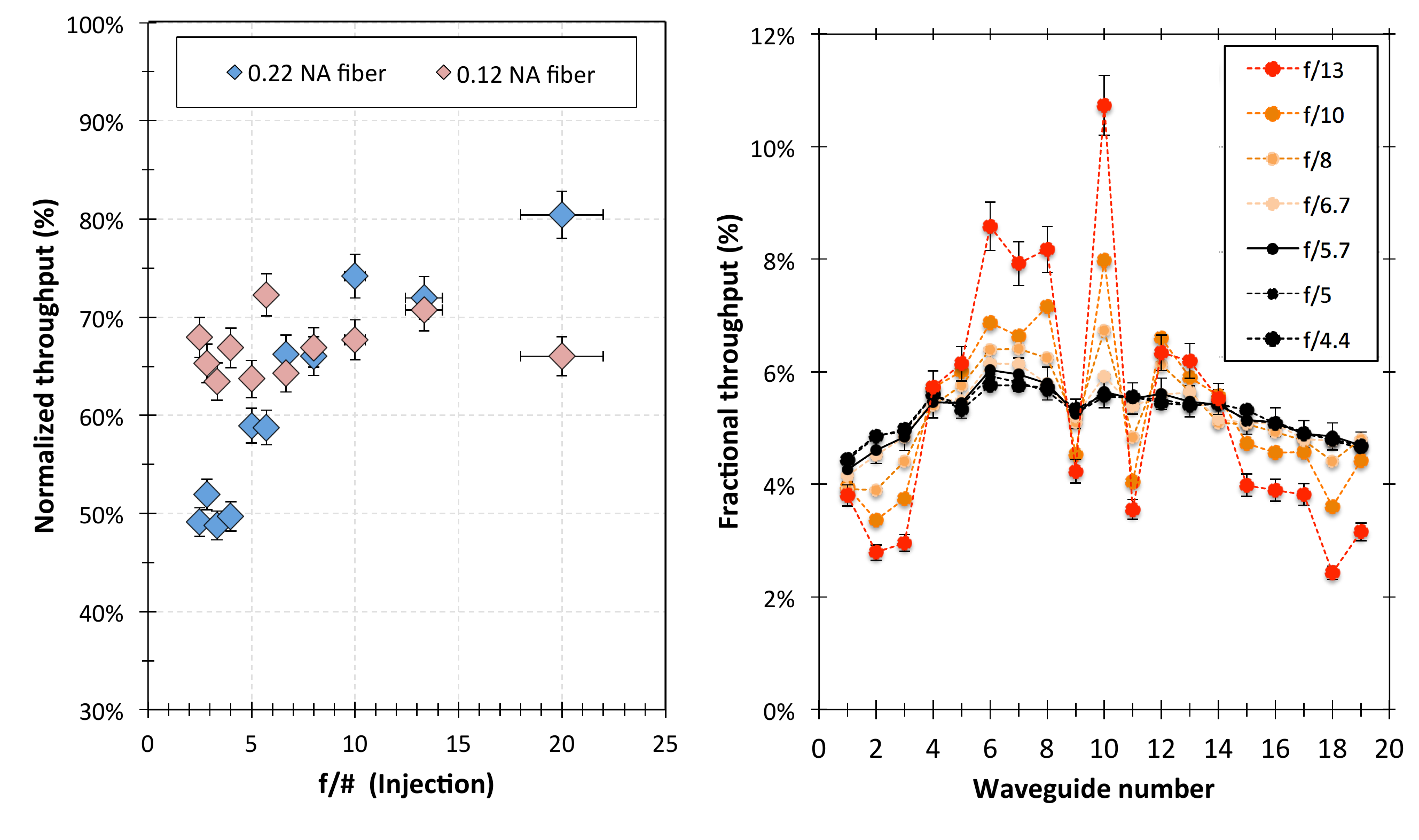}
    \caption{(Left) Throughput of the MMF+IPL device using a low (0.12) and high (0.22) NA fiber as a function of the injected f$/\#$. (Right) Fractional throughput for each SM lantern output as a function of injected f$/\#$. High f$/\#$'s (slower beams) show higher preferential excitement of specific output channels. Low f$/\#$'s show even illumination, and little variation from guide to guide.}
    \label{fig:IPLFnumber}
\end{figure}

The left panel of Fig.~\ref{fig:IPLFnumber} shows the normalized throughput of a broadband light source ($50$~nm at $1550$~nm) as a function of f$/\#$ for two different fiber types, a 0.12 and 0.22~NA fiber with 50~\textmu m core sizes. The low NA fiber (which supports 24 modes compared to the 45 of the 0.22~NA fiber) was used, as it better matched the number of modes supported by the IPL device (19 modes), which had an acceptance cutoff at f$/8$. In this regard, we mean that the low NA fiber restricted the total number of modes incident on the input to the IPL, which should have offered better coupling to the device. The normalized throughput was calculated as the ratio of the flux at the output of the IPL device, divided by the flux coming out of the MMF used to deliver the light to the input to the chip. Therefore, the normalized throughput takes into account Fresnel reflection losses at both ends of the IPL chip, absorption losses in the chip ($\sim20\%$ for the $30~$mm chip used), coupling losses between the MMF and MMWG, transition losses in the IPL, and of course, propagation losses.  

It can be seen that the throughput of the IPL device was $\sim70\%$ at all focal ratios in the case of the low NA fiber, while there was a drop in the throughput for the high NA fiber for f$/\#<7$--$8$. The reason for this has to do with the relative spacing between the effective indices' of the modes in the two fibers. The high NA fiber, which supports many more modes, has a reduced spacing between the effective indices' of the modes. This means that light injected into the input will couple into a larger subset of modes, and any small perturbation to the fiber, be it due to bends or applied pressure on the fiber, will cause energy transfer between the modes, effectively creating a path way for light to move to higher order modes. This phenomenon is called focal ratio degradation (FRD)~\cite{FRD}. For f$/\#>4$ (which was the cutoff of the low NA fiber), both fibers should have been able to efficiently transport the beam to the output and deliver the same power to the IPL device. If this is true, one would expect that the IPL, which has a cutoff of $\sim$f$/8$, would have rejected light at faster focal ratios, and there should have been a drop in the throughput below f$/8$ for both fibers. The fact that the throughput for the low NA fiber remained high instead indicates several things. Firstly, the low NA fiber induced less FRD. This is because there is a larger gap between the modes supported by the guide, and it is more difficult for those modes to cross couple. Secondly, the fiber filtered out some of the higher order modes (i.e. they were more lossy) so that an f$/8$ beam was delivered to the IPL. These are both desirable traits for the instrument we wish to implement this in. Of course, injecting into the fiber at larger focal ratios of say f$/\#=10$ is advisable so that light is not lost in the MMF, which would be wasteful. For these reasons, the low NA fiber was used throughout these experiments.

To gain a more complete picture of the effect that f/\# has on the device, the output of each SMWG was measured as the input focal ratio was adjusted, and the results are summarized in right panel of Fig.~\ref{fig:IPLFnumber}. It can be seen that as the f$/\#$ is adjusted, there is a dramatic change in the illumination of the waveguides. In the regime where f$/\#<6$, the output in each waveguide is similar, with a variation of $<25\%$ from the brightest to the faintest guide. In the regime where f$/\#>7$, the output is strongly modulated, and varies by as much as a factor of $6$ between the brightest and dimmest. It is the latter regime that was utilized in all further experiments, as will be outlined. Although a factor of $6$ does not present a serious limitation in regards to dynamic range considerations, in a final spectrograph built with such a device, it is important to take this into account. Some of this difference will be washed out by the changing input illumination of the point-spread function (PSF) at the focus of the telescope. Nevertheless, it is interesting to observe, and as far as we are aware, is the first time such results have been presented.

%%%%%%%	 On-Sky Results %%%%%%%%%
\section{On-sky results: A new type of modal noise}\label{sec:Onskyresults}
The spectrograph was tested during SCExAO commissioning time on the night of $14^{th}$ and $15^{th}$ April $2014$. The target was W Hydrae (spectral type M7.5, H magnitude of $-2.56$). Once light was visible on the detector, the fiber injection position was adjusted to maximize the flux. Data was acquired with an integration time of 10~s. Several hundred frames were collected, as well as corresponding dark frames afterwards. An average dark frame was computed and subtracted from each frame of data. 

\begin{figure}[!b]
    \centering\subfigure{\includegraphics[width=0.3\linewidth]{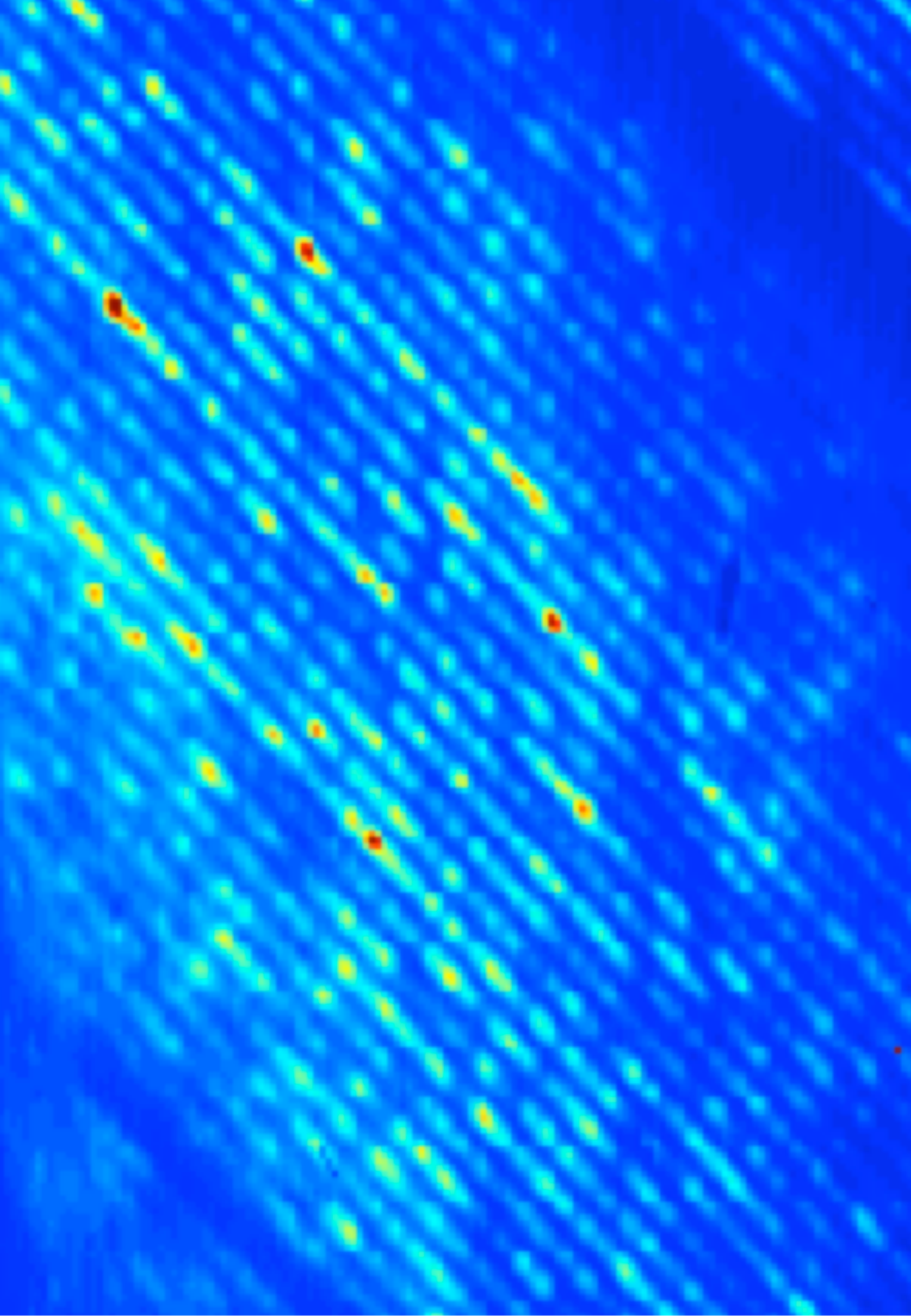}}
	\centering\subfigure{\includegraphics[width=0.3\linewidth]{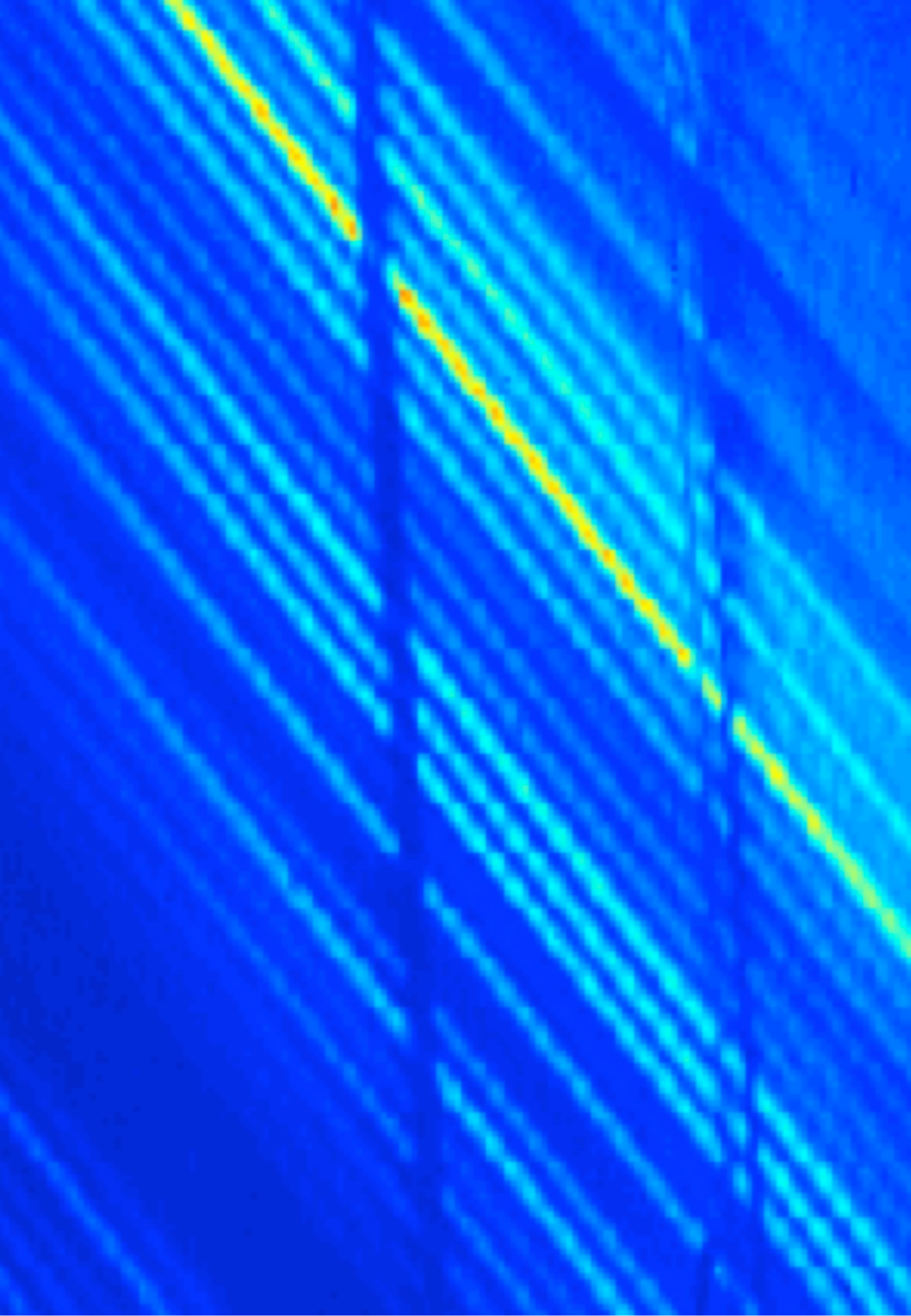}}
    \centering\subfigure{\includegraphics[width=0.3\linewidth]{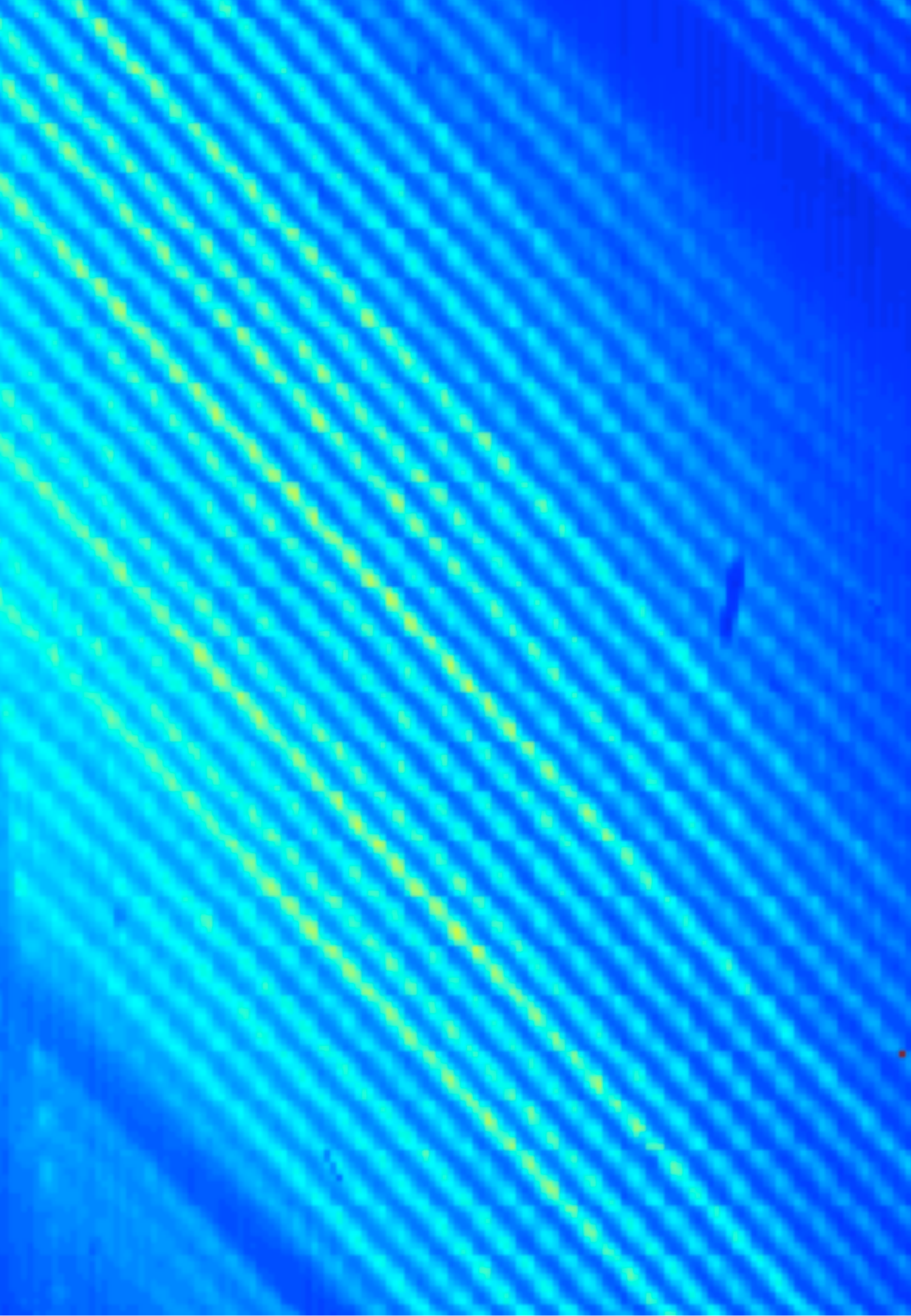}}
    \caption{(Left) A dark-subtracted frame of the spectrograph output showing speckles on the spectral tracks corresponding to the outputs of the PL. The speckles were observed to be time-varying, changing greatly on short time-scales ($<1$~sec), and found to be highly dependent on both the shape of the PSF injected into the MMF, and the movement of the MMF feed. It was visible on all SM spectra, across all grating orders, and found to persist in the co-added final spectra. (Middle) An image showing the AWG output after the MMF was removed, and the light injected directly into the IPL. The vertical stripes in this frame were caused by abrasion on the AWG end face during de-bonding. (Right) An image showing the effect of fiber agitation at high frequencies, compared to the detector exposure time. The speckles were averaged out within each frame. The agitated output is similar to the case of direct injection into the PL (with no input fiber).}
    \label{fig:IPSoutput}
\end{figure}

A portion of a typical image (collected in the laboratory using a broadband light source after the on-sky data was taken) is shown in the left panel of Fig.~\ref{fig:IPSoutput}. It can be seen that there is high frequency speckle pattern, which runs along the spectral tracks of each lantern output channel, also seen in the data collected on-sky. This pattern was not static in the W Hydrae data, but rather changed in both intensity and position between exposures due to remaining higher order modes not corrected by the AO and low order residuals, as well as with environmental changes in the vicinity of the MMF, which induced modal cross-talk. At first glance, it was believed that this was due to the wavelength dependent splitting of the light in the IPL transition region. If this were the case, then by simply co-adding the signal in all the ports of the lantern up, the spectrum of the star could be reconstructed. It was determined that indeed, the channels could not be co-added to reconstruct the spectrum at any point in time, meaning that there must be a wavelength-dependent loss mechanism, preventing all the light making it to the output of the IPL. 

After the on-sky experiments, the MMF was debonded, and light was injected directly into the MMWG of the IPL shown in the middle panel of Fig.~\ref{fig:IPSoutput}. The resulting spectra were smooth and continuous with no evidence of speckles on the tracks. This indicates that it was not a wavelength-dependent loss internal to the IPL chip that creates the modal noise. It must be a result of the interface between the MMF and MMWG, given that it is present when the fiber was used, and not when it was removed. It was believed that this was a new type of loss-mechanism, and associated noise, that was brought about by the variation in intensity of the modes within the MMF, and their subsequent coupling to the modes of the MMWG, which manifests in strong spectral modulations in intensity. This is different to the more traditional modal noise familiar to astronomical spectroscopy, which is a time varying distortion of the spectrograph PSF due to a changing near-field mode distribution of the input MMF feed. The noise manifested, on the other hand, as a highly wavelength dependent loss term (rather than a PSF change), but appeared to be caused by the near-field profile of the MMF as well. This source of noise was completely unexpected in an inherently diffraction-limited instrument. However, this will be common to all astronomical applications where MMFs are used to feed PLs, and warrants further investigation. This kind of noise has not been reported previously in conjunction with these devices.

It is worth noting that this phenomenon was discovered contemporaneously by Spaleniak et al.~\cite{IzaPhD} who used a 35~\textmu m fiber to inject light into a similar ULI fabricated IPL device, to feed the RHEA spectrograph at the Macquarie University Observatory~\cite{feger}.  

%%%%%%%	Source of Noise	%%%%%%%%%
\section{Nature of the modal noise}\label{sec:ModalNoise}

To better understand the nature of this new noise source, and its impact on this type of technology, we carried out a number of tests, including a more focused experimental measurement of the wavelength dependency of the individual photonic components, as well as rigorous beam-propagation simulations of the components used on-sky. This section therefore focuses on firstly presenting the results of the simulations and describing in detail how the modeling was carried out, followed by a comparison to the post-run experimental validation. At the end of the section we discuss the various methods we utilized to remove or suppress the noise.

Before going into the specific details, it is worth elaborating on a few subtleties that were noticed during on-sky testing, which informed our approach for describing and mitigating the noise in this section. When conducting initial alignment of the NIR detector using a broadband source (prior to telescope injection), high injection losses were present, necessitating long exposure times, and the noise was not immediately noticeable. This was because of a combination of two factors: the long exposure time (few seconds) of the detector to overcome the low light levels due to coupling, and the MMF being unintentionally agitated by the environment prior to installation. This washed out the speckles in the output spectra, with the remainder attributed to detector noise. It was only after the installation was complete, with a temporally and spatially stable broadband diffraction-limited light source being injected, and the fiber agitation minimized, that the full impact of the noise became evident.

\subsection{Beam-Propagation Modeling}
To gain a more complete understanding of what is happening to the light inside the photonic components, we modeled the entire light path from the MMF injection onwards using a beam propagation algorithm found in the commercially available package Rsoft BeamPROP. Within the software, we replicated the waveguide structures in $3$-D of the photonic lantern transition, using experimentally measured refractive indices of the SM and MMWG to achieve a more representative index profile. The index profiles of the waveguides was measured using a refractive index profilometer (Rinck Elektronik), with the resulting index cross-section uploaded to the into the software. Thus, the simulation takes into account the complex refractive index structure of the SMWG tracks made by the ULI fabrication method we used. The index measurements, along with a more detailed description of the WG structure and resulting near-field patterns are described in more detail in previous literature on the IPL device \cite{Iza1,jov2012,IzaPhD}. The MMF was simpler to simulate as it had a well-documented step index profile.

\begin{figure}[ht!]
	\centering\includegraphics[width=0.95\linewidth]{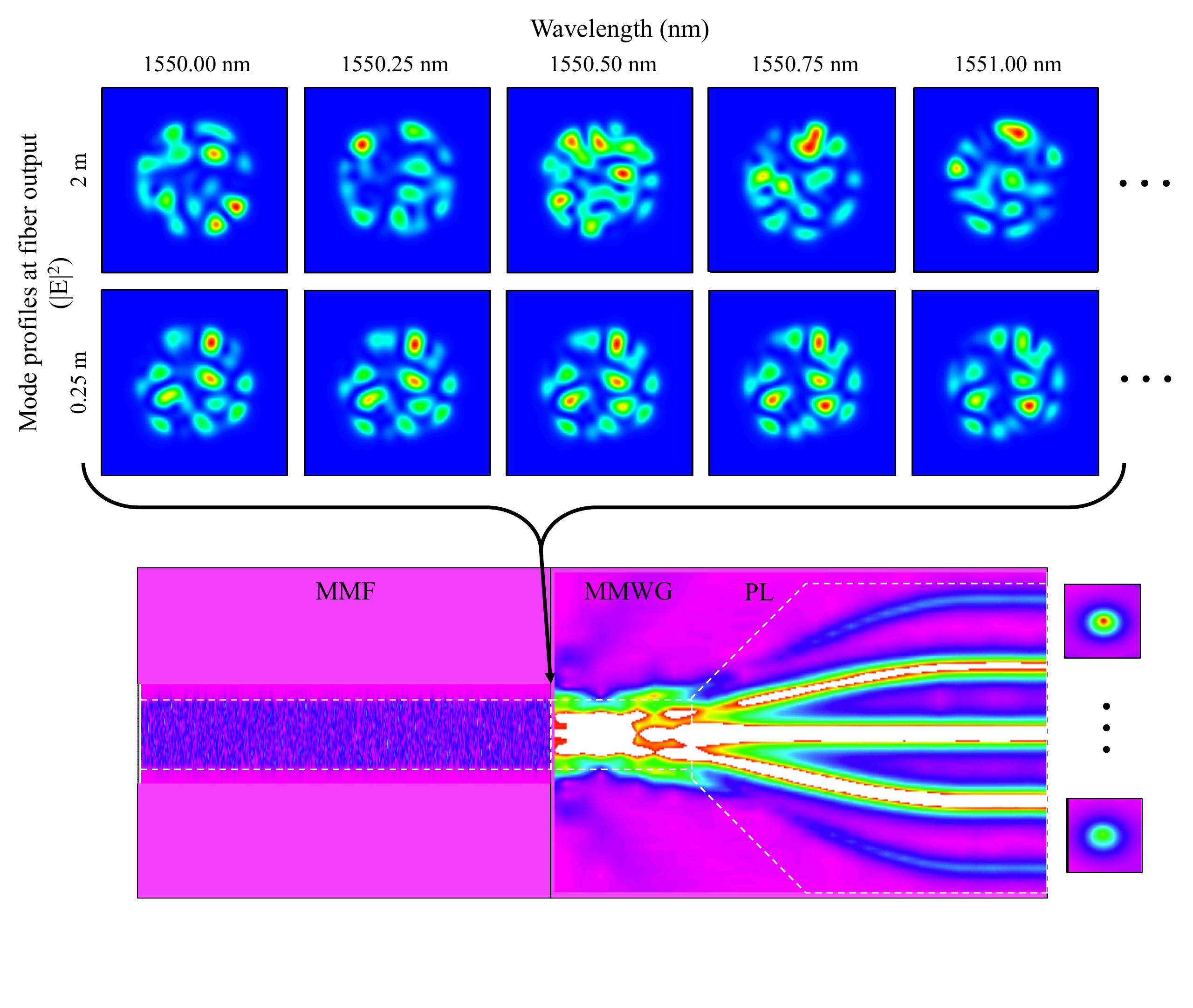}
    \caption{The top series of panels shows results of beam-propagation simulations describing the changes in mode profile as a function of wavelength at the end-face of a 2~m length of MMF, for a fixed input profile. Despite the injected profile having minimal variation in wavelength, the mode profile after propagation in the fiber shows drastic variation at minute changes in wavelength. These are the profiles used to inject into the MMWG section of the IPL. The bottom part is a visualization of the various modeled photonic segments (MMF, MMWG, and PL), made by taking a horizontal slice during a typical simulation, and is for illustrative purposes only. The MMF segment to the left had its aspect ratio adjusted such that it is 2~m horizontally and 250~\textmu m vertically. The evolution of the modes as they propagate is visible as a ripple in the cut. The right segment represents the IPL chip (excluding the remapper transition), and is modeled separately. Because the lantern is 3-dimensional, only 5 of the 19 outputs are visible in this cut. The contrast has been adjusted so the uncoupled light at the MMWG input is visible. The output consists of 19 separate single modes, with two typical outputs shown on the far right.}
    \label{fig:modeprofileSim}
\end{figure}

The modeling was broken up into two distinct steps. Firstly, we propagated monochromatic light through representative MMFs of varying lengths. The wavelength was scanned over a range of $1545$--$1555$~nm in steps of $0.25$~nm (which approximated the on-sky AWG resolution). The launch mode profile was kept identical for all wavelengths and all fiber lengths, and was a 50~\textmu m approximation of a 'seeing limited' spot. At the completion of the simulations, the mode profile data (both the intensity, and the real and complex electric field) was saved for every wavelength. Further, the total encircled power in the fiber was monitored for normalizing the losses. Secondly, the mode profiles were then collated and in turn launched into the MMWG section at the entrance to the IPL as launch modes. The tests were repeated for the entire lantern transition and just the MMWG section to decouple wavelength-dependent losses in the fiber-waveguide interface and the lantern MM-to-SM transition. The light was propagated down the IPL for every wavelength, with the total encircled power at the output of every SMWG recorded. A schematic overview to illustrate this process is shown in Fig.~\ref{fig:modeprofileSim}. 

The upper part of Fig.~\ref{fig:modeprofileSim} shows the output mode profile (intensity) at the end of the MMF after propagation for two different fiber lengths. This illustrates how the fiber near-field is changing as a function of wavelength for the $2$~m fiber (representative of the length used on-sky) and a shorter length of $0.25$~m. For wavelength steps of $0.25$~nm, the near field is dramatically different at each step for the $2$~m length fiber, meaning that at every wavelength element on the final spectrum, a vastly different mode profile is being injected into the IPL, despite the launch mode being perfectly identical for all wavelengths. For the shorter length, the mode profile evolves more slowly as a function of wavelength (i.e. the mode profiles look much more similar). It is important to note that there is no overall drop in power in the MMF itself, but just a change in the distribution within the guide. This is a key part in visualizing the ultimate cause of the noise.  

\begin{figure}[ht!]
	\centering\includegraphics[width=0.99\linewidth]{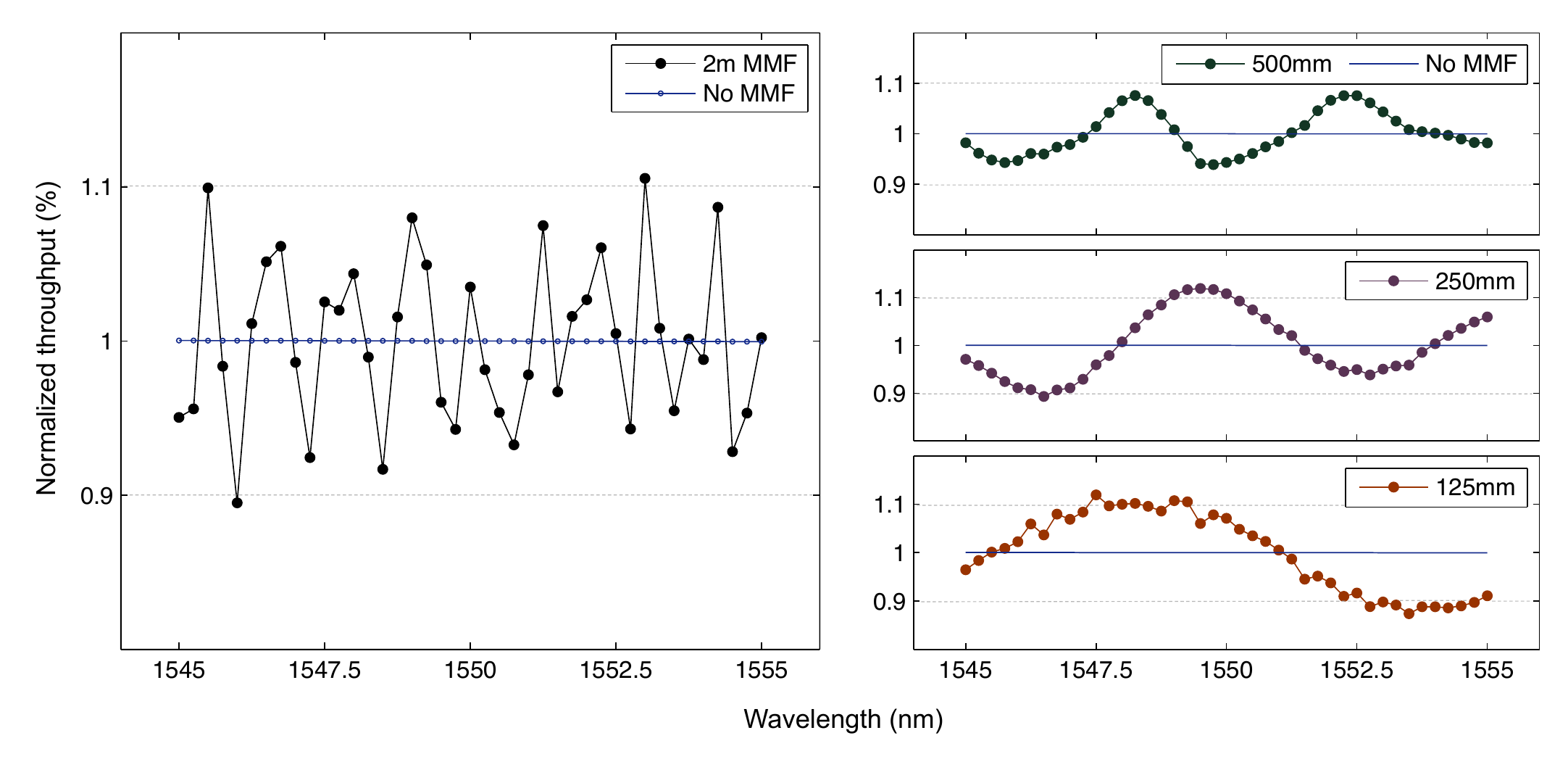}
    \caption{Results from beam propagation simulations demonstrating fluctuations in the throughput of the MMWG as a function of wavelength. Each panel compares the throughput of the MMWG when light is injected with various length MMFs, to the case when the beam is directly injected into the MMWG. For a MMF length of (Left) $2$~m,  (Right top) $0.5$~m, (Right middle) $0.25$~m and (Right bottom) $0.125$~m.}
    \label{fig:MMWGMod}
\end{figure}

The total encircled power measured at the end of the MMWG section (pre-lantern transition), after being injected from a MMF, is shown in Fig.~\ref{fig:MMWGMod}, as a function of wavelength. The left panel shows the case when a $2$~m length of fiber is used, while the right panel shows the case for shorter lengths of MMF. For reference all panels show the result in the case where the original injection mode profile (that was injected into the MMF input) is instead injected directly into the MMWG with no fiber, showing no discernible wavelength dependence at all. The graphs have been normalized such that the mean level of throughput across the spectrum has been removed to emphasize the wavelength dependent coupling loss, which is in the order of $\pm10\%$ in amplitude. 

The results in Fig.~\ref{fig:MMWGMod} provide a few insights into the cause of the observed noise. Firstly, and most importantly, the losses between the MMF and MMWG are shown to fluctuate by around $20\%$ peak-to-peak as a function of wavelength, for all fiber lengths. The simulations did not take into account bends or fiber motion, which would only further exacerbate these spectral modulations. To demonstrate further that this is due to a wavelength-dependent modal-mismatch loss between the MMF and MMWG, the right panel of Fig.~\ref{fig:MMWGMod} provides the results of the losses arising from shorter MMF lengths ($0.125$--$0.5$~m). When the fiber is shorter, the change in the output mode profile as a function of wavelength is much smaller, and gradual enough to be observed by our wavelength sampling. Because the mode profile morphs in shape as a function of wavelength, it moves in and out of ideal matching to the MMWG's acceptance profile, thus modulating the total power coupled into the waveguide. Because for shorter fiber lengths the mode is evolving more gradually, the throughput is modulated across the spectrum with a correspondingly lower frequency. However, the amplitude of modulation is almost constant for all fiber lengths. This indicates that in the regime where the MMF was removed, the wavelength dependent loss should have the same amplitude but the spatial frequency of this would be so low that for all intents and purposes, it would simply be seen as very simple instrument response function which could be removed, and over the bandwidth used in the Fig., would be a flat line as seen. As the length increases, the change in mode profile becomes much more drastic as a function of wavelength, to the point where it becomes uncalibratable. 

Another important insight is that because this is a loss at the MMWG section prior to the MM-SM conversion, the power in the SM outputs of the lantern will no longer add to equal the light coupled into the MMF at the telescope injection (as the loss is internal), and hence cannot be normalized to telescope coupling by any reasonable means. Further, when temporal variation is added (by a changing MMF launch mode, or environmental changes in the MMF) these losses, in the order of 20$\%$, become impossible to predict.

\subsection{Comparison to Experimental Results}
To compare the modeling results presented in this section, we conducted experiments to measure the photonic components under slightly modified conditions to those outlined in Section~\ref{sec:devicecharacterizationsetup}. Firstly, the injection was readjusted to mimic a diffraction-limited Airy pattern at the MMF input. Secondly, the light source was changed to a tunable (with $0.01$~nm wavelength increments) NIR laser, allowing us to measure the throughput as a function of wavelength accurately (analogous to the simulations). Thirdly, because we wished to measure the contribution from each component individually, each component was de-bonded before the next measurement, making this procedure destructive. 

\begin{figure}[ht!]
	\centering\includegraphics[width=0.9\linewidth]{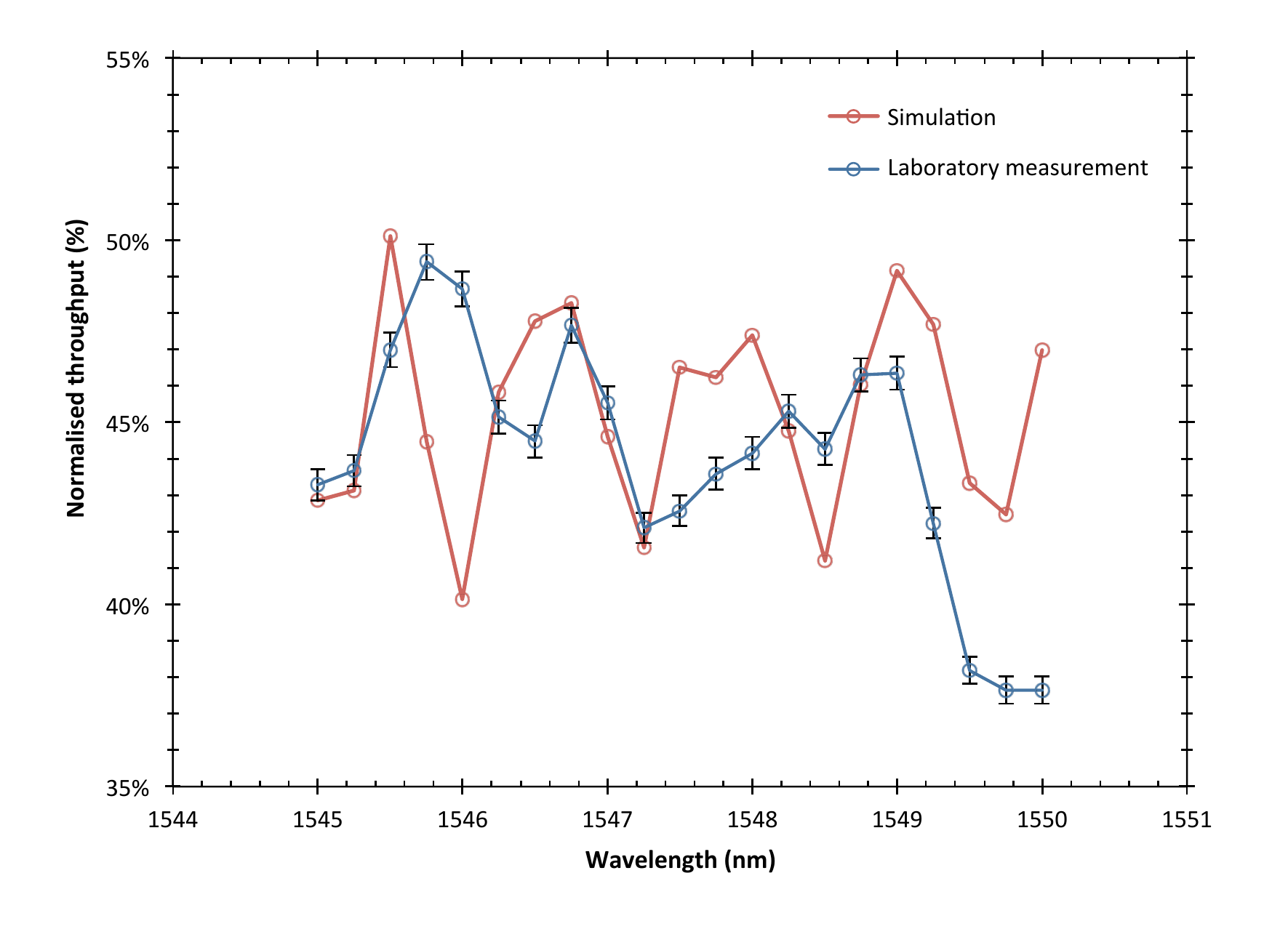}
    \caption{The experimentally measured throughput as a function of wavelength of the MMF bonded to the IPL (AWG removed), and the beam propagation simulation results of the same components.}
    \label{fig:LabNoiseBoth}
\end{figure}

With the AWG output already presented in the on-sky section above, Fig.~\ref{fig:LabNoiseBoth} shows the measured throughput of the MMF and IPL under laboratory conditions (blue) and the overlaid results of the modeling (red). It can be seen that the laboratory throughput is modulated as a function of wavelength in a similar manner to that of the simulations. However, it should be emphasized that the experimental setup was unable to accurately measure the near-field modal pattern at the MMF output simultaneously while measuring the throughput. Further, because the near-field pattern is incredibly sensitive to environmental changes near the MMF, the absolute throughputs are difficult to replicate with much accuracy. Thus, the model was unable to perfectly reconstruct the MMF output modal pattern for each measured wavelength increment and the results presented in Fig.~\ref{fig:LabNoiseBoth} are not expected to match exactly (and where they do is purely coincidental). The key part of the result however, is that the amplitude of the loss is the same order of magnitude for both the simulation and measurement. This demonstrates that the noise observed on-sky was indeed likely caused by the wavelength-dependent modal mismatch between the MMF and MMWG identified by the modeling.  

\begin{figure}[ht!]
	\centering\includegraphics[width=0.7\linewidth]{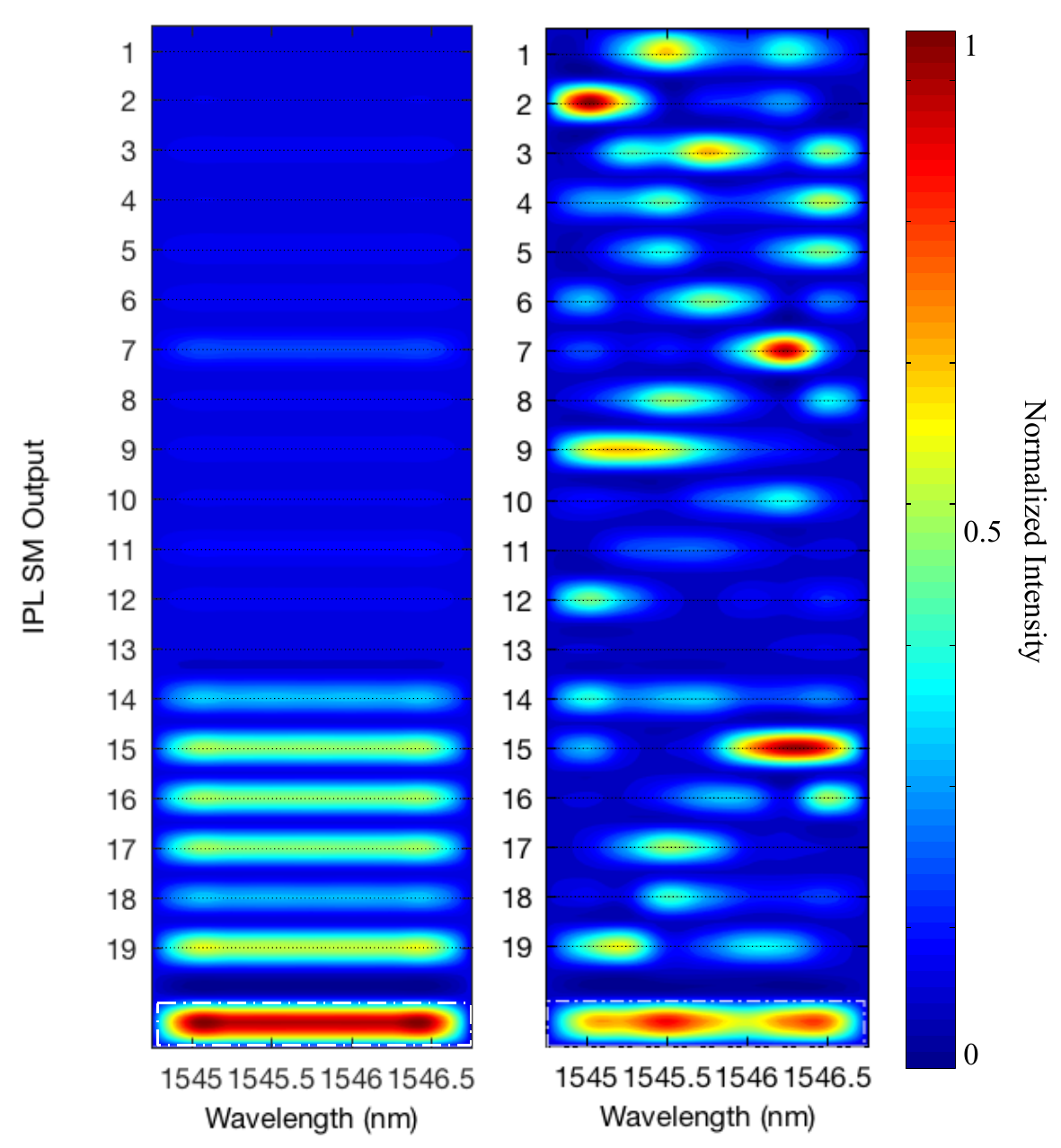}
    \caption{Beam propagation modeling results for the $19$ SM outputs of the IPL as a function of wavelength, when light is injected from (Right) a $2$~m MMF and (Left) no fiber. The individual SM channels are marked with black dotted lines, with the total flux resulting from co-adding all the channels at the bottom marked with a white dotted box. Both graphs have had the simulation output convolved with the measured PSF (output SM mode profile) of the AWG device to obtain similar resolution for comparison.}
    \label{fig:IPLMod}
\end{figure}

Figure~\ref{fig:IPLMod} shows the results of the complete end-to-end model (over a narrower wavelength range) of the full system, from MMF to AWG output for all $19$ SM output channels of the photonic lantern. The left panel shows the output when a diffraction-limited beam is injected directly into the IPL with no fiber. The IPL distributes the flux amongst the individual SMWGs, with some channels having more light in them (in this case $14$--$19$ having the majority), but each channel is constant as a function of wavelength. The sum of all $19$ channels is also uniform and shows no signs of modulation across the spectrum (bottom track). This is the well established typical output of IPL devices.

The right panel however, is of the same system with a $2$~m MMF added with identical launch conditions (hence mimicking our on-sky setup). The modeling results (which were discrete) were convolved with the measured point-spread function of the AWG and cross-dispersion system so as to represent more closely what it would look like on the detector. Here, the speckle is immediately evident in two respects. Firstly, the total co-added flux of all the SM channels (bottom track) has the noise present, and is same order of magnitude as the data presented in Fig.~\ref{fig:MMWGMod} and~\ref{fig:LabNoiseBoth}, showing why the on-sky spectra were not calibratable. Secondly, because the mode profiles entering the IPL are now vastly different as a function of wavelength, the flux-distribution in the SMWG channels also varies as a function of wavelength. Thus, the true nature of the noise is not only a fluctuation in the total throughput of all channels, but a further fluctuation in the SM channels as well. The frequency and intensity of the speckle matches that seen in on-sky results.

\subsection{Removing the Modal Noise}

% Because the wavelength dependent loss is caused by the mismatch between the MMF output near-field pattern and the acceptance of the MMWG, and because the MMF output is very sensitive to both the injection PSF and environmental changes near the fiber, in practice they combine to cause the speckles in the final spectrum to be temporally varying. 

Under typical conditions the speckles in the final spectrum are temporally varying, which arises due to a combination of two effects. Firstly, the wavelength dependent loss is caused by the mismatch between the MMF output near-field pattern and the acceptance of the MMWG, and secondly, the MMF output is very sensitive to both the injection PSF and environmental changes near the fiber. While problematic for calibration, this effect can be exploited to average out the noise by ensuring that the integration time of the detector is orders of magnitude longer than the temporal variance of the output MMF near-field. Essentially this averages out the wavelength dependent losses to a single scalar component. 

Figure~\ref{fig:IPSoutput}~(right) shows the laboratory outcome of the simplest approach, agitating the fiber feed at speeds much faster than the exposure time (kHz agitation frequencies for an exposure time of a second in our case). Because the oscillations in the agitator caused variations in the strain of the fiber, the MMF output modal distribution changed on similar timescales, thus effectively averaging out the speckles during the $1$~s exposure. In essence, this averaging out of the MMF near-field is what traditional fiber agitators are doing for MMF-fed high resolution spectroscopy. It should be noted that this approach averages down the time varying noise amplitude, but does not remove it completely. Further, if agitating the fiber causes higher order modes to be excited that are not able to be collected by the MMWG, the light is lost and decreases the overall throughput. For low resolution applications, the agitation will likely prove adequate, however for high-precision applications, the noise is still a major issue because in this regime long-term stability of the spectra is required. Further, this approach negates one of the benefits of single-mode spectrographs, which under normal conditions have a non-varying PSF and don't require agitation. 

In both this and the first approach for mitigating this effect, the tracks will indeed appear smoother. However, the wavelength-dependent coupling loss will still occur at the MMF/MMWG interface. This will be averaged out to a single DC value, but means that absolute flux calibration will not be possible by either approach.  

An alternative approach, and one that physically removes the noise entirely, is to remove the input fiber feed and inject directly into the MMWG. While this configuration would bring our device more in line with the injection methods of other groups (i.e. injecting directly into the lantern)~\cite{HarrisOnSky}, it removes the advantage of having a fiber feed with the spectrograph assembly to be positioned at the focal plane, which was one of the main drivers for this work. Another recent approach pioneered by MacLachlan et. al.~\cite{MacLachlan} is to revert back to the use of fiber PLs, which have no fiber-WG interface, and create an integrated remapper to reformat the $2$D output array of a multicore-fiber lantern, into a $1$-D array. This ostensibly uses fibers to perform the MM-SM conversion and WGs to remap, much like the back-end of the IPL we presented here. While recent analysis showed that this approach can remove this form of noise~\cite{IzaRemap},  similar effects can arise when trying to form pseudo-slits (slab WGs) at the output. It should also be noted that the noise can be removed by ensuring that the MMWG and MMF have the exact same near-field structure (unlike those presented here), by using more advanced beam-shaping techniques and creating a more uniform MMWG structure. Thus, as the ULI fabrication methods mature and perfect replication of MMF feed becomes achievable, the wavelength-dependent loss at the interface can potentially be entirely eliminated. 

Finally, it should be mentioned that while fiber agitation and direct lantern injection are ways to mitigate this issue, it can be sidestepped entirely by injecting the telescope light directly into an SMF. While this requires high quality AO correction to achieve high coupling efficiencies, it is rapidly becoming a reality~\cite{jov2017}. In fact, this ultimately was the approach we adopted for the IPS prototype testing, and is presented in Jovanovic et. al.~\cite{jov2017AWG} in the same special issue.

%%%%%%%  	Summary		%%%%%%%%%
\section{Conclusion}\label{sec:summary}
We have demonstrated the realization of an all-photonic device capable of both MM-to-SM conversion and spectral dispersion on an 8-m class telescope with efficient coupling. During on-sky testing, a new and previously unreported form of wavelength dependent loss, which manifests as modal noise, was discovered, that made spectral extraction and calibration extremely difficult. The source of the loss and associated noise was traced to a wavelength-dependent loss mechanism between the MMF near-field pattern and the modal acceptance profile of the MMWG structure inside the IPL. Modeling of the photonic components corroborated this hypothesis, by replicating the wavelength-dependent loss and demonstrated an identical effect on the final spectral output. We outlined that this could be mitigated by directly injecting into the IPL. As photonic components become implemented into astronomical instrumentation on greater scales, studies such as this identify potential pitfalls that might otherwise trap unsuspecting instrument designers. When working in the astrophotonic regime, ignoring the modal nature of light comes at a cost.

\section*{Funding}
Japan Society for the Promotion of Research ($23340051$, $26220704$, $23103002$). Australian Research Council Centre of Excellence for Ultrahigh bandwidth Devices for Optical Systems (CE$110001018$).

\section*{Acknowledgments} This work was supported by the Astrobiology Center (ABC) of the National Institutes of Natural Sciences, Japan, the directors contingency fund at Subaru Telescope and the OptoFab node of the Australian National Fabrication Facility. S. G. acknowledges funding by the Macquarie University Research Fellowship scheme. The authors wish to recognize and acknowledge the very significant cultural role and reverence that the summit of Maunakea has always had within the indigenous Hawaiian community. We are most fortunate to have the opportunity to conduct observations from this mountain.

\end{document}